# A Low-Profile, Self-Contained System for Atmospheric Monitoring and Mid-flight Collection of Viable Microbiological Samples at High Altitude

Caitlyn A. K. Singam[1]
*University of Maryland, College Park, MD, 20740, USA*

The prevalence of bacteria in the atmosphere has been well established in relevant literature, suggesting that airborne bacteria can influence atmospheric characteristics including the development of clouds. Studies have also demonstrated that the atmospheric biological profile is influenced by the underlying terrestrial biomes. An understanding of the complex interplay of factors that can influence the atmospheric biological profile, not to mention developing a biological census of the atmosphere, requires a cost-effective experimental system capable of generating reproducible results with reliable data. However, as has been demonstrated by payloads launched by space agencies such as NASA and JAXA, these payloads are both complex and cost prohibitive. This paper discusses the design and implementation of a biologically oriented experimental payload for high-altitude ballooning that is within the means of most student-run experimental programs. The payload highlighted in this presentation, PHANTOM (Probe for High Altitude Numeration and Tracking of Microorganisms, which has the goal of capturing aerial microorganisms at multiple altitudes in order to characterize the biological composition of the upper atmosphere), has undergone a number of successful flight trials, and serves to highlight the feasibility and utility of interdisciplinary projects between aerospace and the biological sciences.

## I. Introduction

Although the literature contains documented evidence of microbial spores in atmospheric dust dating back more than a century – with one of the earliest references to such samples appearing in the 1830 treatise by Berkeley on cryptogamic botany [1] – there has been a distinct dearth of documentation on the characteristics, spread, and effect of atmospheric microbiota in relation to their aerial environment. This is in no small part driven by the longstanding assumption that such organisms were insignificant players in the atmosphere, incapable of effecting any sort of significant change due to the transience in the atmosphere and inability to survive at high altitudes that was presumed at the time. However, in recent years, bacteria have been discovered at altitudes of 8 to 15 km, with further evidence suggesting the possibility of permanent microbial colonies in the upper troposphere and above. A 1935 study by Meier [2] reported evidence of microbial life at even higher altitudes in the range of 70 to 80km. These stable populations of high-altitude bacteria (henceforth referred to as 'aetherophiles') have been uncharacterized for the most part due to the limited scope of previous sampling of the microbiome. Most prior research has focused on low-altitude bacterial populations, whose numbers and compositions are constantly in flux and easily influenced by changes in weather and temperature.

Aerial micro-organisms in particular have garnered interest from biologists due to the extreme environment and unusual conditions presented by high altitude, and the possibility of new, hereto undocumented adaptations to those conditions being discovered in micro-organisms that have either adapted to withstand or thrive in high altitude conditions. The existence and characteristics of terrestrial and aquatic extremophiles have already been well documented in the literature [3],[4],[5], as well the characteristics of a diverse collection of micro-organisms with dedicated adaptations to ionizing radiation [6], thermal shock [7], and other such conditions [8],[9], and would provide a well-documented point of comparison for any newfound microbial strains.

---
[1] Undergraduate student, Department of Biology, University of Maryland, College, AIAA Student Member

The presence of aerial microbes at high altitude also is a subject of interest for those in the aerospace and meteorology fields as well, due to the fact that the presence of microbiota in large bodies of water (which are subject to evaporation) [10] and in the atmosphere has been established as being an important component of climate regulation and the development of local weather patterns. Additionally, the problems inherent in designing a biological sampling system capable of withstanding the conditions of the upper atmosphere while simultaneously maintaining sample integrity, viability, and stability are non-trivial and pose a challenge from a biological and engineering perspective [11].

Prior work has suggested the presence of relatively high densities of microbial colonies in the troposphere and at higher altitudes [12], though isolating samples collected at such altitudes from contaminating terrestrial micro-organisms has proven challenging [13]. As a result, the literature has reported inconsistent results as to the exact character and species [14],[15] of the organisms present at various altitudes, though studies have indeed been consisting in documenting the presence of some sort of microbial life at altitudes exceeding 10 km above sea level. DeLeon-Rodriguez et. al., in a widely publicized 2013 study in the Proceedings of the National Academy of Sciences [16], utilized planes as a means of facilitating air sampling by carrying a system on-board capable of filtering air at a 20 L/min sample rate via a mechanically actuated air sample system. Their results, after performing air sample collection at an altitude of 20 km, yielded a total of 75 colony forming units, which evidence indicated were terrestrial in origin and likely transient denizens rather than permanent aerial colonies. Griffin's 2003 report [17] on a similar attempt at biological sampling at high altitude also reported the presence of culturable microbes (albeit a more varied collection of species), which was consistent with larger, and slightly different set of species identified from samples collected in a subsequent 2018 study [18].

Based on this information, it is difficult to predict what exactly the results of any biological sample collection-based experiment at high altitude might be in terms of the specific species or strains that might be captured, especially given the lack of consistent results regarding the identity of microbes present at such altitudes – or, for that matter, any sort of significant body of results/large sample size of obtained via the same methodology. The dearth of such large sets of reproducible data in the literature is undoubtedly attributable in no small part to the immense cost of the experimental setups employed in most published studies; a some of the more notable studies in the literature (including the aforementioned ones) utilized plane-based collection systems [19] that allowed for the investigators to minimize the risk of contamination (by ensuring careful handling of the sampling equipment) by virtue of bypassing the twin constraints of weight and financial burden that tend to afflict projects spearheaded by smaller research groups. Indeed, the issue of financial constraints is severe enough to have primarily limited research attempts into aerobiological profiling solely to groups led by a small pool of researchers backed by large national space agencies.

The system described herein, the Probe for the High Altitude Numeration and Tracking of Micro-organisms (PHANTOM) is a high altitude balloon (HAB) payload designed with the intention of facilitating the identification of micro-organismal species across various altitudes through a live-capture methodology implemented in a cost-effective system accessible to a significantly larger scientific audience than previously developed methodologies. For such a design to be effective, it must adhere to a strict set of design parameters and requirements. Specifically, it must be capable of collecting viable samples from the atmosphere at distinct altitudes with minimal contamination from terrestrial sources, while simultaneously acquiring environmental data that can provide the context for the collected samples as well as insight into the environmental conditions that may have driven the development of any characteristics or deviations from terrestrial species or other documented strains.

## II. Design parameters

High altitude balloon (HAB) flights provides a versatile and accessible means of studying the atmosphere and near-space environment, particularly through a biological lens. In terms of accessibility, the cost of a single high altitude balloon launch can be as low as $1000 – well within the means of small research groups and some student-run teams – depending on the size of balloon used and the amount of lift needed for the payload string. HAB flights also do not rely on any active propulsion that might cause excessive disturbances in the pre-existing atmospheric biome (e.g. turbulence), a key characteristic for observational payloads with a biological focus.

HABs have the additional benefit of being highly modularized and easily customizable. Unlike with other aerial vehicles, it is fairly trivial to alter a number of flight-critical variables within certain tolerances (e.g. target burst altitude, lift capacity, etc.) on the launchpad of a HAB flight if necessary, or to interchange various flight-critical parts with spares in the case of an unexpected failure (e.g. inadvertent and/or premature termination of the balloon). This is well-suited for biological experimental systems, where one may need to alter the speed at which the balloon ascends or the altitude at which the balloon bursts in order to ensure the optimum payload exposure to conditions of experimental interest.

A standard HAB flight using a 1600 g latex balloon, filled with 13 $m^3$ of helium and with a payload string approximately 5 kg in weight can last one to three hours on average, and can reach 25 km to over 30 km in altitude. The bulk of the flight time is taken up by the ascent phase, during which the balloon climbs at a rate of around 5 m/s (if one assumes around 2 kg of free lift). This provides adequate time for a biological payload to receive exposure to near-space conditions and is a sufficient velocity to allow for airflow needed to collect samples during ascent. As a passive vehicle, the balloon is also highly subject to the types of powerful air currents that move large quantities of airborne bacteria, and thus has the capability of following the direction of microbial migration during its journey to the stratosphere. HABs like the one described herein are capable of traveling an average of 72 km downrange from the initial launch site, and their flight patterns describe sprawling arcs that can cover a wide geographical area and a number of different terrestrial biomes.

However, there are a number of constraints and challenges that have to be considered when designing a biological experimentation system for use on a HAB flight. Most notable among these constraints are weight and dimensions; given the importance of attaining high altitudes for most flights, payload designers are required to keep their payload weight to an absolute minimum, and are thereby somewhat limited in their choices of construction materials and parts. Although foam core board is a structural material of choice for payload exteriors, using easily degradable material is restrictive for biological payloads that require repeated sterilization with liquid disinfectants. Thin sheets of aluminum alloy or other such lightweight metals are a viable option, but are more difficult to work with and take time to machine. Liquids of any sort are also not permissible in payloads due to the sub-freezing temperatures experienced during flight and the potential for spillage or container rupture.

The potential for mechanical damage must also be taken into account during payload design. During flights, payloads experience dramatic shifts in environmental conditions while in the air, and need the capability to endure a gamut of environmental conditions that might otherwise stress the external structure of the payload and threaten sensitive items inside the payload. Pressure changes, for instance, can place stresses on the box that can cause it to either explode or implode if it is not properly vented, which is a potential issue for a biological payload that needs to minimize or entirely eliminate outside contamination. In addition, payloads need to be able to withstand high-velocity winds during descent along with what is usually a moderately violent landing. After burst, HABs can reach speeds of close to 70 m/s prior to parachute deployment, and even subsequent to parachute deployment can impact the ground or other terrestrially-based obstacle at a speed of 5-10 m/s [20]. The balloon's trajectory and flight path is highly variable and outside an experimenter's control once the HAB leaves the launch pad, making obstacle avoidance a particularly difficult challenge.

Arguably the most imperative requirement of any payload, however, is to adhere to financial constraints. The financial burden associated with either buying or fabricating parts for a payload is not insignificant, especially for payloads with more complex designs or housing sophisticated equipment. This particularly affects biological payloads, where standard experimental equipment such as the biological air samplers recommended by International Organization for Standardization standards (ISO 14698) [21] can cost upwards of $1500 – more than the operational cost of some HAB launches. In order to keep a biological payload within the means of a student-operated ballooning group, such costly equipment has to instead be substituted with innovative and unconventional solutions that harness the unique environmental conditions associated with HAB flights in an experimentally viable manner.

Even with the challenges that face payloads designed for HAB flights, it is nonetheless possible to design, construct, and test biological experimental systems capable of obtaining scientifically viable data and samples. PHANTOM (Probe for High Altitude Numeration and Tracking of Microorganisms), the payload described herein, is a testament to the viability of such a system.

## III. Payload implementation

PHANTOM was designed with the intention of facilitating the reconnaissance and identification of micro-organismal species across various altitudes in a cost-effective, lightweight, and spatially conservative system. With long-term use, such a system is theoretically capable of collecting a sufficiently large data set with enough samples from different points within any one geographical region and over a sufficient span of time to allow for the density of different aetherophilic species to be mapped across various altitudes, independent of launch location.

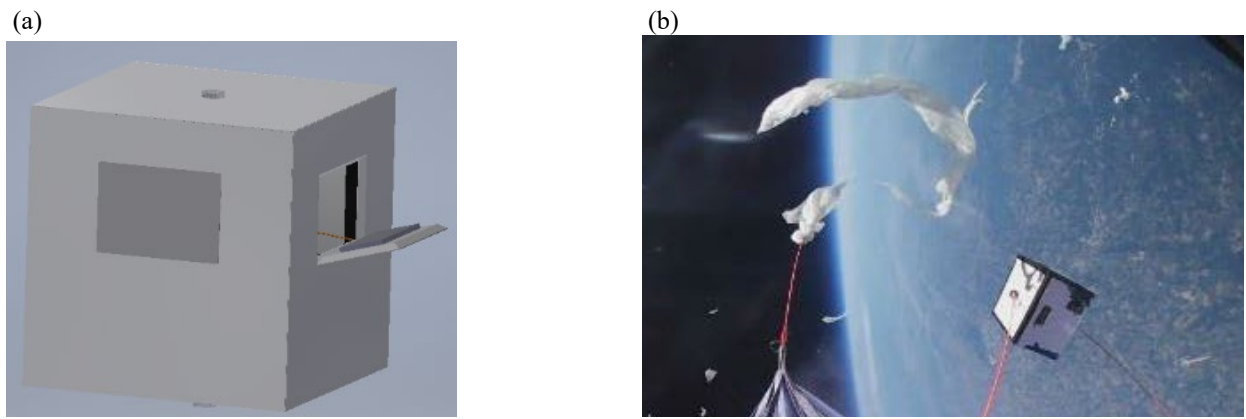

**Figure 1. The external structure of PHANTOM.** *(a) A CAD drawing showing the external structure of PHANTOM, with one of its sampling mechanisms in its actuated form. (b) A photograph of PHANTOM during a high altitude balloon flight, where it is visible next to the remnants of the burst balloon.*

The proof of concept payload utilizes a conventional cuboidal external structure favored by a number of high altitude ballooning groups and is constructed out of foam-core material (i.e. 3 mm of polystyrene foam laminated between outer facings of clay-coated paper) lined with 12.7 μm thick polyvinyl chloride (PVC) plastic. The foam core material was chosen due to its low density, low cost, and ready availability, while the PVC lining allows for gas exchange between the payload and outside air and easy pre-flight sterilization while simultaneously preventing the entry of micro-organisms into the sterile payload chamber during flight by virtue of its thinness and material properties. It is separated into an upper region (containing four sample compartments) and a lower electronics bay, is ventilated in both regions. The external structure of the payload (illustrated via a CAD model in Fig. 1a, and in its implemented form in Fig. 1b) minimizes the amount of external features in order to reduce weight and maximizes payload safety, as it ensures that no components are mounted in a position where they are capable of being detached from the payload mid-flight. Instead, the electronics are mounted internally, and are accessible via a side door. In order to prevent the electronic components from moving during flight, fabric-backed adhesive tape is used to secure them in place in the electronics bay. Most of the electronics are encased in the same 12 mm insulative polyethylene foam which lines the inside of the electronics chamber in order to prevent cold-induced device failure mid-flight.

Each of the sample collectors included in the payload operate using a servo motor attached to a door in the side of the payload that facilitates air exchange with the external air. The door is opened when the on-board barometric altimeter (which doubles as a pressure/temperature sensor) indicates that the payload is at one of the pre-determined sample altitudes, at which point the servo actuates to let the door open in a type of drawbridge-style mechanism (visible in Fig. 1, on the actuated sample collector). The servos are powered through a motor driver connected to an Arduino Uno, a readily available, low-cost microprocessor known for its open-source documentation and compatibility with a number of different commercial sensors.

## IV. Testing methodology and results

### A. Bench tests

In order to verify the functionality of the sample collection activation and actuation mechanism, the payload was ground tested on numerous occasions by transporting the payload to areas of different elevations to simulate ascent (albeit on a smaller scale), and observing whether the doors were able to open at an intended pre-set altitude. PHANTOM consistently demonstrated functionality of its sample collection mechanisms (the fact that the tests were done on the ground rather than in-flight allowed for direct visual confirmation of the actuation of each sample collector), thus corroborating the data collected from prior flight tests that indicated that the doors opened successfully.

## B. Flight tests

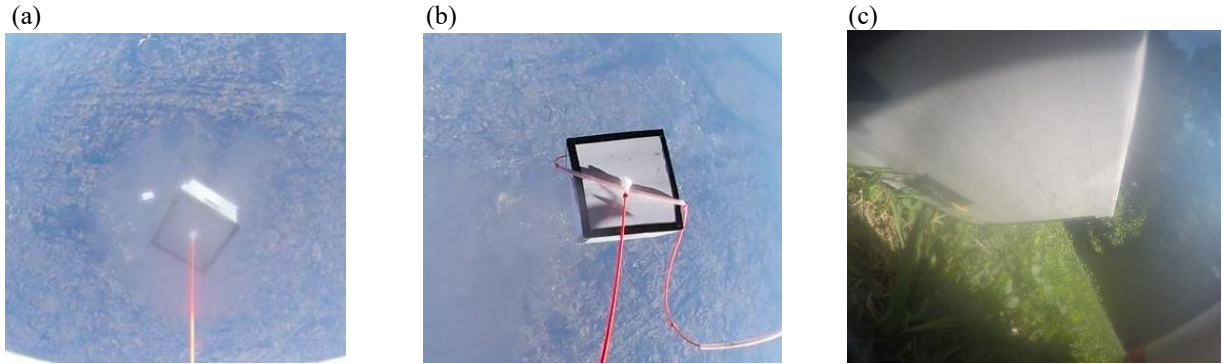

**Figure 2. Photographs from the PHANTOM flight tests.** *(a) Photographic evidence of PHANTOM remaining attached to the payload string under strong winds during the March 31, 2018 flight; circled in red is another payload that flew alongside PHANTOM, falling off the payload string. (b) PHANTOM was buffeted by the remnants of two payloads which fell off the string mid-flight (March 31, 2018); the remnants of both payloads (plastic tubing) is visible in the photograph. (c) Photographic evidence showing PHANTOM intact and safely attached to the payload string immediately post-flight (June 30, 2018) and after impacting the ground on landing.*

The current iteration of PHANTOM, featuring the drawbridge-style sampling mechanism, flew on two high altitude balloon flights (which were launched on March 31, 2018, and June 30, 2018, respectively). During the flights, the payload experienced drastic changes in environmental conditions, inclusive of temperature drops, rapid pressure changes during balloon ascent and descent, high radiation levels (which have previously resulted in the structural compromise of exposed 3D printed plastic components on payloads that have flown alongside PHANTOM, and can thus be considered a risk to payloads), and mechanical stress, which it needed to withstand in order to remain intact and safely attached to the payload string. On both flights, PHANTOM was able to continuously collect environmental data via its onboard sensors, and demonstrate the integrity of its attachment system by remaining securely attached to the payload string. Other payloads did not fare as well, and some payloads that have exhibited structural integrity issues and design flaws pre-flight have fallen off the payload string on the very same flights where PHANTOM survived without damage (Fig. 2a and 2b). Furthermore, PHANTOM also remained consistently unharmed post-landing (Fig. 2c) and the state of the payload post-flight did not necessitate any repairs in order to return the payload to a flight-ready state, which suggests that the proof-of-concept prototype is in fact viably resilient and capable of withstanding flight conditions.

Post-flight data analysis, of both the data collected by the on-board sensors and visual inspection of the payload sample collection mechanisms, appeared to indicate that the payload collected viable data (Fig. 3a and Fig. 3b are examples of some of the graphs generated from the collected data) and also successfully actuated its sample collectors. Although the opening of the sample collectors was not readily visible from videos and photographs taken during the flight, due to a combination of the swinging of the payload in and out of the frame of the camera during flight, and the relatively small size of the sample collectors, the payload data recording system provided an alternate means of assessing the success of the collection actuation mechanism. Verification of the sample collectors opening was done by examining the payload's on-board data log (using Boolean variables that recorded the status of each sample collector) that all four collectors were triggered at the appropriate altitudes and received enough voltage to actuate. Furthermore, a rudimentary "dead man's switch" mechanism located inside each of the sample collectors on the proof-of-concept flights (comprised of an ultrafine string which broke when the collector actuated and the drawbridge mechanism opened) indicated that, on both flights of the current PHANTOM prototype, all four collectors actuated during flight.

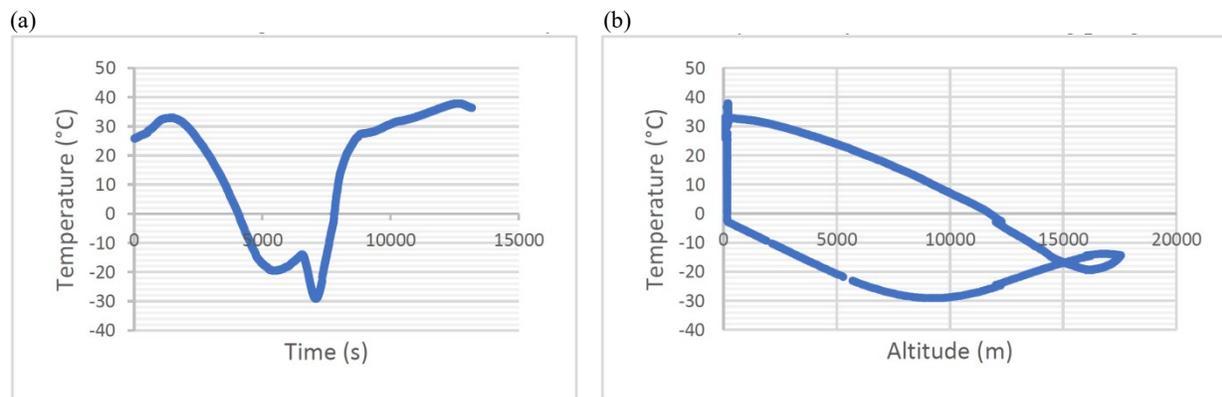

**Figure 3. Flight data from PHANTOM during a high altitude balloon flight.** *Both graphs illustrate that the data obtained is within range for what is experienced by high altitude balloon payloads; the temperature data in (a) has clear enough resolution for the temperature changes in the tropopause to be clearly visible. The temperature vs. altitude graph in (b) similarly follows the expected pattern.*

Previous iterations of PHANTOM have also maintained a perfect safety record on multiple high altitude balloon flights; given that the nature of the design process inherently means that many of the design processes and principles underlying even the earliest versions of the payload have carried over to the current version of the payload (particularly with regard to attachment systems and an avoidance of the conventional paperclip-based system), it is not unexpected that the payload was able to withstand high altitude flight conditions with ease.

Test runs were also done with previous iterations of PHANTOM to test sterility as well. For a sterility test, the payload was sterilized pre-flight using a benzalkonium chloride and ethanol-based disinfectant and were primed with sterile lab-grade filter paper on the inside of doors in order to simulate the setup of an actual experimental flight, were shown to maintain sterility when the doors remained closed (this was done by leaving one chamber closed throughout the entire flight as a control) and to be successful in collecting viable samples (predominantly composed of *E. coli*) when exposed to the air at altitudes of and 5 km and 10 km.

## V. Conclusion

The success of the PHANTOM design in withstanding the vicissitudes of HAB flight and of its preliminary sterility tests indicate the feasibility of low-cost live-capture systems for atmospheric profiling that are cost-effective and that have a minimal impact on the aerial microbiome that the system in question is seeking to document. Additionally, the data collected by payloads such as PHANTOM, which furthers understanding how biological material may survive, thrive and be transported via the upper atmosphere has myriad applications, inclusive of national security implications (with regards to the dispersion of aerosolized biological agents) in addition to meteorological and medical significance. The design for PHANTOM itself and the feasibility of sample collection in harsh environments also has potential applications for the design of biologically oriented missions with the aim of identifying possible signs of life on exoplanets.